\documentclass[conference]{IEEEtran}
\usepackage{enumitem}
\usepackage{bm}
\usepackage{amsmath,amssymb,amsfonts}
\usepackage{algorithmic}
\usepackage{graphicx}
\usepackage{textcomp}
\usepackage{fancyhdr}
\usepackage[hyphens]{url}
\usepackage{comment}
\usepackage{bm}

\usepackage[switch]{lineno}

\usepackage[ruled,linesnumbered,noend]{algorithm2e}
\SetKwComment{Comment}{$\triangleright$\ }{}
\SetAlFnt{\small}

\SetCommentSty{mycommfont}
\usepackage[usenames,dvipsnames,table,xcdraw]{xcolor}
\usepackage{mathtools}
\usepackage{diagbox}
\usepackage{subcaption}

\usepackage[sort,nocompress]{cite}
\usepackage[bookmarks=true,breaklinks=true,letterpaper=true,colorlinks,linkcolor=black,citecolor=blue,urlcolor=black]{hyperref}

\usepackage{enumitem}

\usepackage{booktabs} 
\usepackage{float}
\usepackage{makecell}
\usepackage{color}
\usepackage{tikz}

\newcommand{\insertFigure}[2]{
    \begin{figure}[t]
\setlength{\abovecaptionskip}{0pt}
\setlength{\belowcaptionskip}{0pt}
        \centering
        \includegraphics[width=\linewidth]{figures/#1.pdf}
	\vspace{-2mm}
        \caption{\small #2}
	\vspace{-2mm}
        \label{fig:#1}
    \end{figure}
}

\newcommand{\insertWideFigure}[2]{

    \begin{figure*}[h]
\setlength{\abovecaptionskip}{0pt}
\setlength{\belowcaptionskip}{0pt}
        \centering
        \includegraphics[width=\textwidth]{figures/#1.pdf}
	\vspace{-2mm}
        \caption{\small #2}
	\vspace{-2mm}
        \label{fig:#1}
    \end{figure*}

}

\newcommand{\squishlist}{
 \begin{list}{$\bullet$}
  { \setlength{\itemsep}{0pt}
     \setlength{\parsep}{3pt}
     \setlength{\topsep}{3pt}
     \setlength{\partopsep}{0pt}
     \setlength{\leftmargin}{1.5em}
     \setlength{\labelwidth}{1em}
     \setlength{\labelsep}{0.5em} } }

\newcommand{\squishlisttwo}{
 \begin{list}{$\bullet$}
  { \setlength{\itemsep}{0pt}
     \setlength{\parsep}{0pt}
    \setlength{\topsep}{0pt}
    \setlength{\partopsep}{0pt}
    \setlength{\leftmargin}{2em}
    \setlength{\labelwidth}{1.5em}
    \setlength{\labelsep}{0.5em} } }

\newcommand{\squishend}{
  \end{list}  }

\newcommand*\circled[1]{\tikz[baseline=(char.base)]{
            \node[shape=circle,fill=darkgray,inner sep=0.5pt] (char) {\textcolor{white}{#1}};}}

\newcommand{\hardtaco}{\textsc{Hard TACO}\xspace}

\newcommand{\aespa}{\textsc{AESPA}\xspace}

\def\BibTeX{{\rm B\kern-.05em{\sc i\kern-.025em b}\kern-.08em
    T\kern-.1667em\lower.7ex\hbox{E}\kern-.125emX}}

\title{Enabling Flexibility for Sparse Tensor Acceleration via Heterogeneity} 
\author{
    \IEEEauthorblockN{Eric Qin\IEEEauthorrefmark{1}, Raveesh Garg\IEEEauthorrefmark{1}, Abhimanyu Bambhaniya\IEEEauthorrefmark{1}, Michael Pellauer\IEEEauthorrefmark{2}, Angshuman Parashar\IEEEauthorrefmark{2}, \\ Sivasankaran Rajamanickam\IEEEauthorrefmark{3}, Cong Hao\IEEEauthorrefmark{1} and Tushar Krishna\IEEEauthorrefmark{1}
    \IEEEauthorblockA{\IEEEauthorrefmark{1}\textit{Georgia Tech}, Atlanta, USA.
    \{ecqin,raveesh.g\}@gatech.edu, callie.hao@gatech.edu, tushar@ece.gatech.edu}
    \IEEEauthorblockA{\IEEEauthorrefmark{2}\textit{NVIDIA}, Westford, USA.
    \{mpellauer, aparashar\}@nvidia.com}
    }
    \IEEEauthorblockA{\IEEEauthorrefmark{3}\textit{Sandia National Labs}, Albuquerque, USA.
    srajama@sandia.gov}
    }
\begin{document}
\maketitle
\pagestyle{plain}

\begin{abstract}
Recently, numerous sparse hardware accelerators for Deep Neural Networks (DNNs), Graph Neural Networks (GNNs), and scientific computing applications have been proposed. A common characteristic among all of these accelerators is that they target tensor algebra (typically matrix multiplications); yet dozens of new accelerators are proposed for every new application. The motivation is that the size and sparsity of the workloads heavily influence which architecture is best for memory and computation efficiency.
To satisfy the growing demand of efficient computations across a spectrum of workloads on large data-centers, we propose deploying a flexible `heterogeneous' accelerator, which contains many `sub-accelerators' (smaller specialized accelerators) working together.
To this end, we propose: (1) \hardtaco, a quick and productive C++ to RTL design flow to generate many types of sub-accelerators for sparse and dense computations for fair design-space exploration, (2) \aespa, \underline{a} h\underline{e}terogeneous \underline{sp}arse \underline{a}ccelerator design template constructed with the sub-accelerators generated from \hardtaco, and (3) a suite of scheduling strategies to map tensor kernels onto heterogeneous sparse accelerators with high efficiency and utilization. \aespa with optimized scheduling achieves 1.96$\times$ higher performance, and 7.9$\times$ better energy delay product (EDP) than a Homogeneous EIE-like accelerator with our diverse workload suite.

\end{abstract}

\section{Introduction}

Large datacenters are expected to compute a wide variety of workloads such as deep neural networks, graph neural networks, and scientific computing \cite{reddi2020mlperf, mattson2019mlperf,DLRM19,fey2019fast,li2019pasta}. These workloads utilize tensors (a matrix is a two-dimensional tensor, and a vector is a one-dimensional tensor) of different dimension sizes and sparsity characteristics. For example, matrix dimension sizes span from single digits to millions while matrix sparsity spans from $\sim10^{-5}$\% dense to fully dense \cite{suitesparse}. The vast amount of workloads has led to many accelerator architecture proposals, as they achieve higher throughput than CPUs, and higher energy efficiency than GPUs \cite{tpu-isca, rocki2020fast, shao2019simba}. There are numerous types of sparse accelerators because they typically target a specific application, which often have tensors with similar dimensions, sparsity, and structure. 

Many of these accelerators are rigid 
in both the dataflow choice 
as well as the sparsity format employed. This makes them perform extremely well for certain workloads, but poorly for other workloads. 
For instance, a systolic array is most energy-efficient for dense computations, but not for workloads of high unstructured sparsity. ExTensor \cite{extensor}, on the other hand, performs well for workloads of high unstructured sparsity, but not for dense computations due to its sparse controller overhead.   
Large datacenters require flexibility, as in they must have the compute and memory resources to perform all current and future workloads efficiently. 
\textbf{To address this challenge, we propose a new \textit{heterogeneous} sparse accelerator and scheduling techniques to enable high efficiency across a diverse set of workloads}. To realize this architecture, we identify three challenges that we address in this work.

\insertFigure{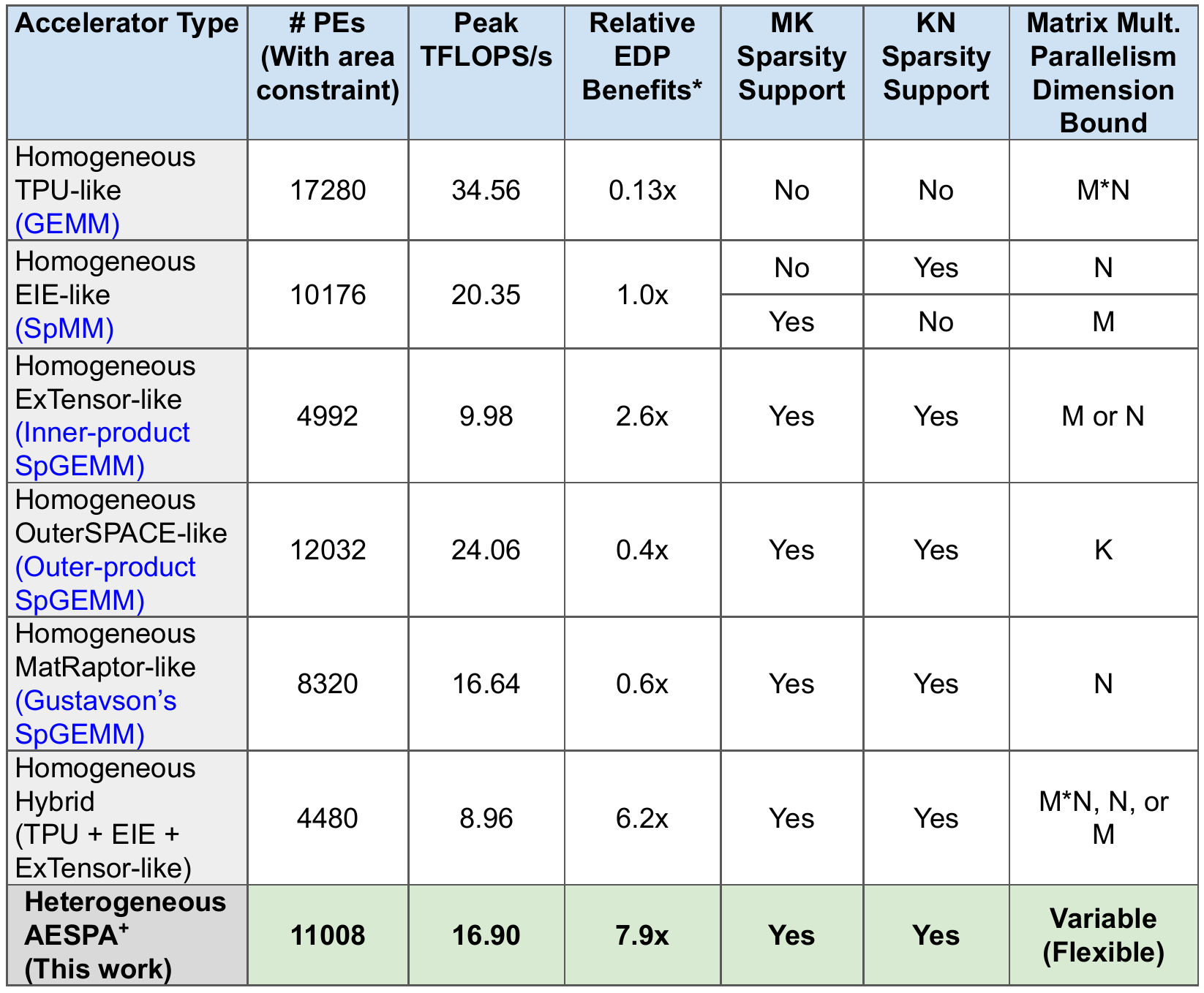}{Design characteristics of homogeneous sparse accelerators (consisting of one type of sub-accelerator) versus heterogeneous sparse accelerators (consisting of 2+ types of sub-accelerators). Matrix \textbf{A} is M $\times$ K and Matrix \textbf{B} is K $\times$ N. \vspace{1mm}}

\textbf{Challenge 1: Building Blocks.} The first challenge is to identify the right sparse sub-accelerator building blocks.
Recently, numerous sparse architectures have been proposed for different applications \cite{eie, srivastava2020tensaurus, eyeriss2, parashar2017scnn, outerspace, zhang2020sparch, extensor, sigma, zhang2021gamma, srivastava2020matraptor, yan2020hygcn, geng2019awb, geng2019awb}. There seems to be dozens of new sparse matrix multiplication accelerators proposed for every new application. This is a growing problem because there are many sparse accelerators that are similar. For example, both MatRaptor \cite{srivastava2020matraptor} and Gamma \cite{zhang2021gamma} utilize Gustavson’s Algorithm. 
Detailed explanation is in Section~\ref{sec:hardtaco}. 

To address this challenge, we turn to TACO, a popular sparse tensor compiler \cite{kjolstad2017tensor,tacoonline,chou2018format}.
Many of the sub-accelerators can be described at an algorithmic level using insights from TACO. `Dense' sub-accelerators operate on uncompressed input tensor operands, while `sparse' sub-accelerators operate on compressed input tensor operands. The compression format combination of the input tensors influence the type of dataflow the sparse sub-accelerator utilizes, which is discussed in Section~\ref{sec:background}. 
\textbf{\textit{We propose a new sparse sub-accelerator design methodology named \hardtaco that can generate the RTL for different distinct sub-accelerators for fair performance, power, and area estimates.}}
\hardtaco contains a hardware fine-tuning stage that adds pragmas on top of the TACO output kernel C++ code (shown in Fig~\ref{fig:taco-output-new}). The updated code then goes to a HLS tool to generate functional sparse accelerators and controllers. Similar high productivity hardware generation tools include MAGNet \cite{venkatesan2019magnet} and Deepburning-GL \cite{liang2020deepburning}, but they cannot generate all types of sparse sub-accelerators that \hardtaco can. 

\textbf{Challenge 2: Hardware Design-space Exploration.}  The second challenge relates to identifying the appropriate design-point using the sub-accelerator building blocks and contrast it against homogeneous alternatives.
From a physical die perspective, different sub-accelerators consume different area and power overhead. Under a given area constraint, sub-accelerators with smaller PEs can achieve higher TFlops/second than sub-accelerator with larger PEs which have significantly more control overhead. On the other hand sub-accelerator with larger PEs due to control logic can handle sparsity more efficiently. \textit{Additionally, to build a heterogeneous sparse accelerator, it is important to allocate the right amount of compute and memory resources.} Some sub-accelerators require more memory accesses than others, and often lead to lower operational intensity. Fig~\ref{fig:intro-table} shows different types of homogeneous accelerators (consisting of one type of sub-accelerator). A homogeneous TPU-like accelerator \cite{tpu-isca} computes GEMM operations (dense matrix $\times$ dense matrix), and a homogeneous EIE-like accelerator \cite{eie} computes SpMM operations (sparse matrix $\times$ dense matrix or dense matrix $\times$ sparse matrix). Homogeneous ExTensor \cite{extensor}, OuterSPACE \cite{outerspace}, and MatRaptor-like \cite{srivastava2020matraptor} accelerators all compute SpGEMM operations (sparse matrix $\times$ sparse matrix); utilizing inner product, outer product and Gustavson’s algorithm respectively. The different microarchitecture and sparse controller for each sub-accelerator results in different processing element (PE) sizes. With a normalized area constraint, a Homogeneous TPU-like accelerator can achieve 3.4$\times$ higher peak 
TFLOPS/second. Additionally, the type of dataflow used for these accelerator can limit which the amount of parallelism achieved within the accelerator. The
OuterSPACE accelerator is bounded by the workload's K dimension; therefore, there will be underutilization if the K dimension is smaller than the number of OuterSPACE PEs available. 
Fig~\ref{fig:intro-table} also presents homogeneous hybrid accelerators. These type of accelerators contain the necessary sparse controller to support multiple types of operations at the cost of lower max TFLOPS/s. 

\textbf{\textit{To address this challenge, we propose \aespa (\underline{a} h\underline{e}terogeneous \underline{sp}arse \underline{a}ccelerator) that efficiently interconnect many different sub-accelerators and scratchpad memories}}. 
\aespa provides flexibility (variable parallelism dimension bounds) through heterogeneity by having many types of sub-accelerators while achieving higher TFLOPS/s than the homogeneous hybrid approach across a diverse set of workloads.

\textbf{Challenge 3: Scheduling/Mapping.}  The third challenge, is to determine what scheduling strategies provide the best performance, energy efficiency, and utilization.
This is actually the first work (to the best of our knowledge) to unravel the challenges of scheduling on heterogeneous sparse sub-accelerators while previous works look solely into heterogeneous dense sub-accelerators \cite{herald, prema, planaria, baek-isca}. 

\textbf{\textit{To address this challenge, we propose various scheduling strategies aimed at utilizing all sub-accelerators.}} Specifically, we look into strategies that can (1) partition a single tensor kernel to utilize \textit{all} sub-accelerators, and/or (2) partition independent tensor kernels across diverse sub-accelerators for multi-tenancy.
Different sparse sub-accelerators may require input tensors to be of a certain format before computation (discussed in Section~\ref{sec:hardtaco}), which we address by placing custom hardware format converters into \aespa.

\vspace{1mm}
\noindent  \textbf{\textit{In summary, the key contributions of this paper are:}}
\begin{itemize}

\item We propose a new class of sparse tensor accelerators that leverages the idea of heterogeneous sub-accelerators, each optimized for a specific sparse dataflow (and corresponding compression format).

\item We develop \hardtaco, a productive sparse and dense sub-accelerator generation design flow for quick performance, area, and power analysis. (Refer to Section~\ref{sec:hardtaco}.) 

\item We design \aespa, a heterogeneous sparse accelerator that stitches different types of sub-accelerators together. (Refer to Section~\ref{sec:aespa}.) 

\item We propose various scheduling strategies for heterogeneous sparse accelerators. (Refer to Section~\ref{sec:scheduling}.) 

\item Our findings show that \aespa with optimized scheduling achieves 1.96$\times$ higher performance, and 7.9$\times$ better energy delay product (EDP) than a Homogeneous EIE-like accelerator with our diverse workload suite. 

\end{itemize}

To the best of our knowledge, this is the first work to analyze how to build a heterogeneous sparse accelerator with various sub-accelerator types, and the first work to propose scheduling policies for such accelerators.
\section{Background}
\label{sec:background}

This section first introduces the main compression formats used in this work. We use a taxonomy to express matrix multiplication algorithms and and sparse accelerator dataflows based on the operands' compression format. 




\subsection{Sparse Compression Formats}
There are numerous types of compression formats, both structured and unstructured. 
For this taxonomy, we focus on unstructured compression formats used directly for computation. 
We refer to it as \textit{compute compression format (CCF)} throughout the rest of the paper. To express CSR or CSC (shown in the top section of Fig~\ref{fig:sub-accelerators}), we follow a naming scheme inspired by TACO \cite{chou2018format} and ExTensor \cite{extensor}. Each dimension of a matrix (two-dimensional tensor) can be uncompressed or compressed. With this method, CSR and CSC can be thought of as the same format, but compressed in a different mode orientation \cite{li2018hicoo}. 

Following the M $\times$ N $\times$ K matrix multiplication convention, Matrix \textbf{A} has dimensions M $\times$ K and Matrix \textbf{B} has dimensions K $\times$ N. If Matrix \textbf{A} is stored in CSR format, then there is a row pointer for every row and a column index for every nonzero value. We represent CSR by $U_MC_K$ with `U' meaning \textit{`uncompressed'} and `C' meaning \textit{`compressed'}. The subscript variables represent the dimensions of the matrix. Dimension M is considered uncompressed as CSR follows a row-major ordering, and each row location must be specified by the row pointer. Alternatively, if Matrix \textbf{A} is stored in CSC format, we represent it by $U_KC_M$. If Matrix \textbf{B} is stored fully uncompressed (dense), it is $U_KU_N$.

\insertFigure{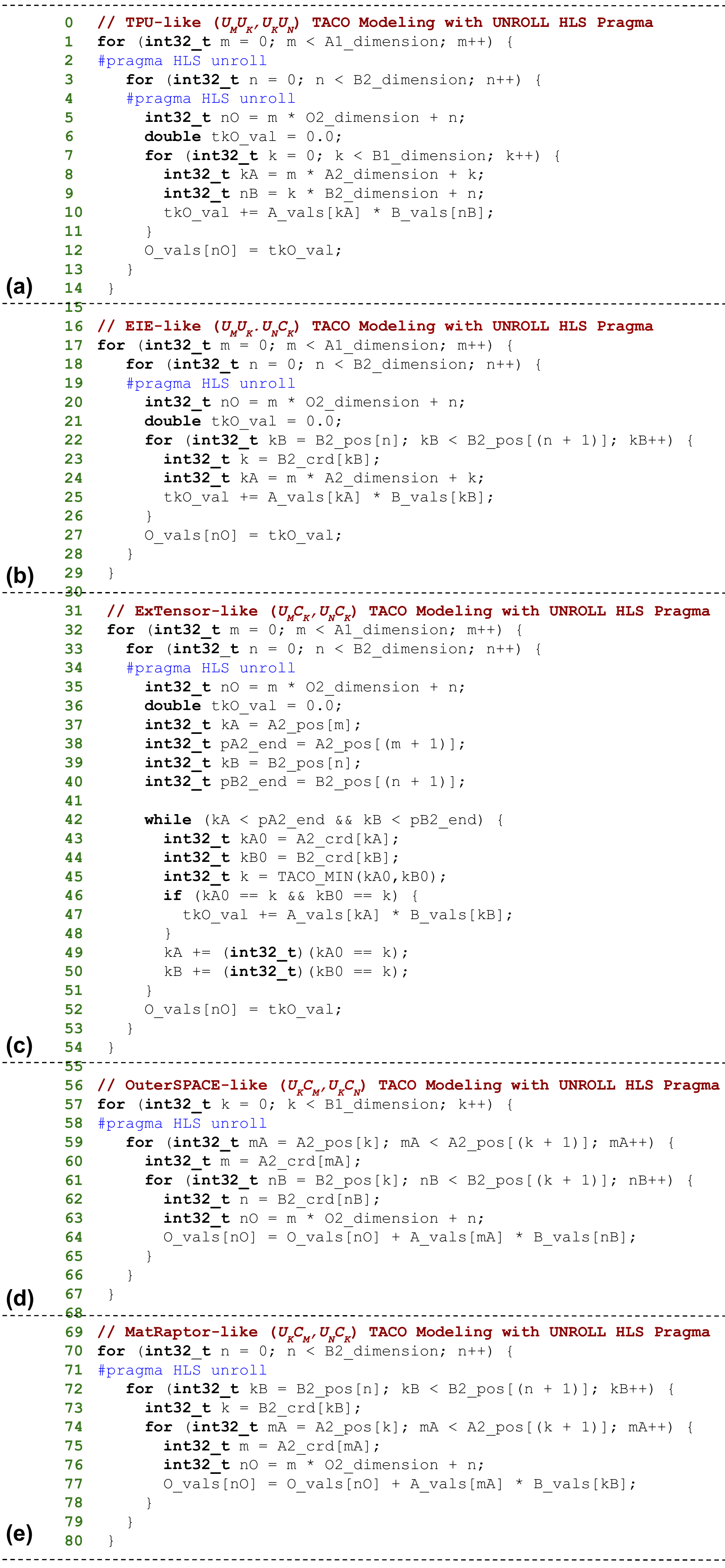}{TACO generated matrix multiplication kernels with UNROLL HLS Pragma for different sub-accelerator types. Operand casting code omitted for brevity. } 

\insertWideFigure{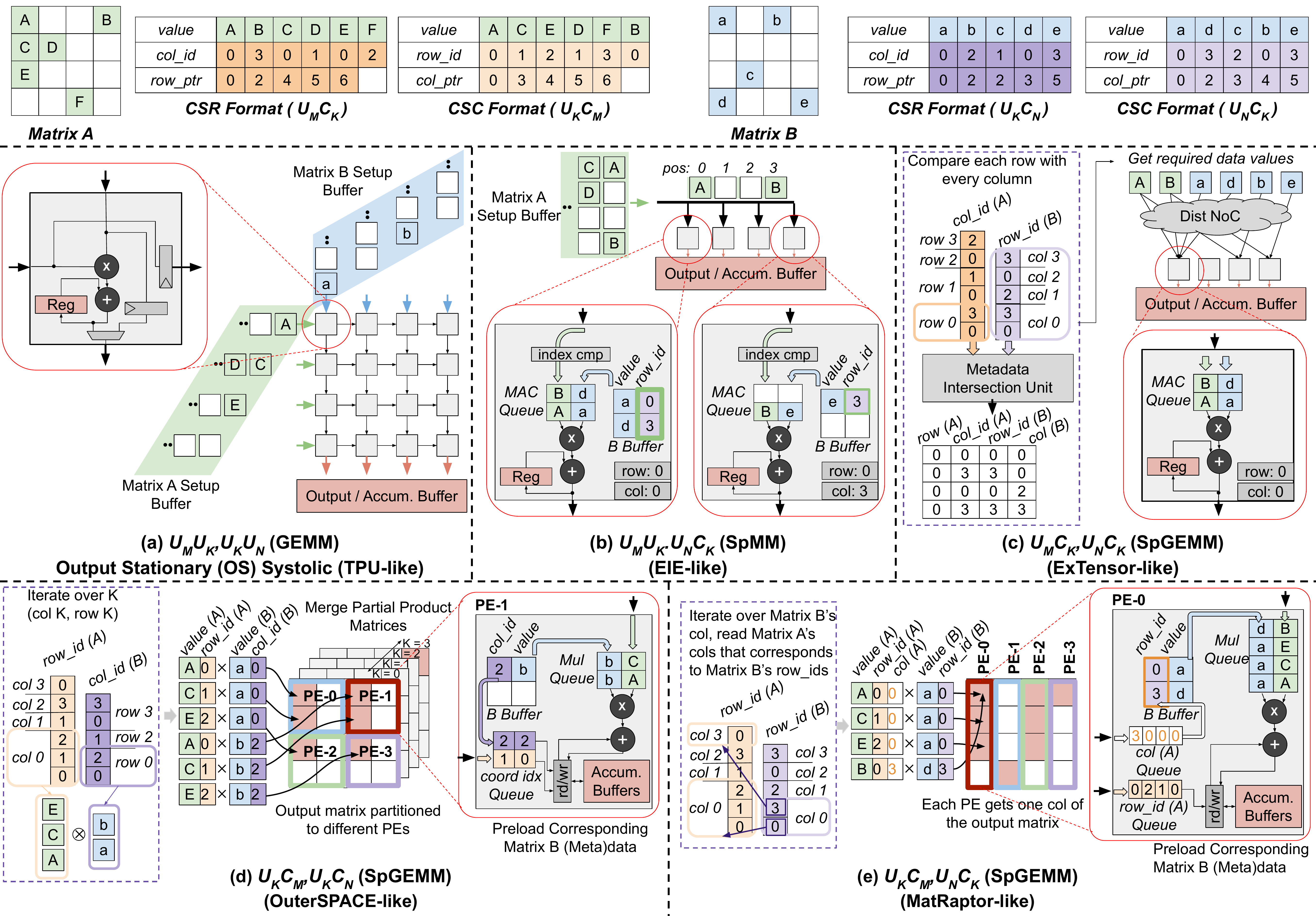}{Examples showing different MatMul accelerator classes. The top row shows two input matrices along with their compressed equivalents. $value$ represents the nonzero value. $col\_id$ and $row\_id$ represent the coordinate (crd) of the nonzero value. $row\_ptr$ and $col\_ptr$ represent the position (pos) where a row or column begins. Inspired from state-of-the-art sparse accelerators, (a-e) presents how different Compute Compression Formats (CCFs) combinations can be mapped onto hardware. \vspace{3mm}}

\subsection{Proposed Taxonomy with TACO}
\label{sec:background_taxonomy}


Real world tensors are often sparse, which lead to ineffectual computations (e.g. multiplications with zero-valued operands) on dense hardware. CCFs enable algorithms that can skip these computations by utilizing bookkeeping metadata to indicate where the nonzero elements are located. Each tensor operand has its own compression format. Throughout the rest of the paper, we only read the CCFs in concordant fashion (following the same mode orientation). We propose a taxonomy template for matrix multiplications that follows:

\begin{center}
$(\bm{A,B}):format(Tensor A),format(Tensor B)$
\end{center}

To illustrate the taxonomy template and how ineffectual computations are skipped, we use TACO  (a tensor algebra compiler for kernel generation) to generate five different matrix multiplication kernels shown in Fig~\ref{fig:taco-output-new}.  For example, Fig~\ref{fig:taco-output-new}c shows a SpGEMM kernel with Matrix \textbf{A} compressed in $U_MC_K$ and Matrix \textbf{B} compressed in $U_NC_K$. 

Using our template, we refer to this as:

\begin{center}

$(\bm{A,B}):U_MC_K,U_NC_K$

\end{center}

Fig~\ref{fig:taco-output-new}a shows a TACO generated kernel with both matrices computed uncompressed ($U_MU_K,U_KU_N$). All loops depend on the dimensions of the GEMM operation. Since there is no way to distinguish whether the input matrices' elements are zero-valued or nonzero-valued, all computations are scheduled regardless of sparsity. Fig~\ref{fig:taco-output-new}b shows a TACO kernel with the CCF combination of $U_MU_K,U_NC_K$. Matrix \textbf{B} is compressed with only nonzero values along the K dimensions, as shown in line 22. Matrix \textbf{A} is uncompressed, so a zero-value (from Matrix \textbf{A}) $\times$ nonzero-value (from Matrix \textbf{B}) computation can still occur. To completely eliminate any ineffectual computations from zero-value multiplications, both matrices must be compressed. Fig~\ref{fig:taco-output-new}c shows a TACO kernel with the CCF combination of $U_MC_K,U_NC_K$.  Both matrices hold position indices (also commonly referred to coordinates) of the nonzero values along the K dimension, and both indices must match (refer to line 46) to indicate a valid nonzero computation. Fig~\ref{fig:taco-output-new}d shows a TACO kernel with the CCF combination of $U_KC_M,U_KC_N$. Line 57 shows that the kernel is iterating over the uncompressed K dimension, while line 59 and line 61 show that it is iterating over compressed M and N dimension respectively. Different CCF combinations will generate different TACO outputs. Although the compression format of the output matrix (\textbf{O}) may vary, we default it to fully uncompressed (dense) for the rest of the paper, and can be hidden within the taxonomy unless compressed otherwise. The TACO kernels with different CCFs can be correlated to the computation behaviors of dense and sparse accelerators. The next section will describe more in detail on how Fig~\ref{fig:taco-output-new}a,b,c,d,e correlate to TPU \cite{tpu-isca}, EIE \cite{eie}, ExTensor \cite{extensor}, OuterSPACE \cite{outerspace}, and MatRaptor \cite{srivastava2020matraptor} respectively.

\section{Hard TACO: Sparse Sub-Accelerators Generator}
\label{sec:hardtaco}

This section first introduces different sparse sub-accelerator microarchitectures, each utilizing a TACO kernel discussed in Section~\ref{sec:background}. Then, this section proposes \hardtaco for productive generation of hardware for fast performance, power, and area estimates. 

\subsection{Classifying Custom Sparse Accelerators}
\label{sec:sparseclass}

In this subsection, we use the TACO based taxonomy to classify state-of-the-art sparse accelerators. The examples following show exactly how the compression formats are used for sparse computations, and which matrix dimension(s) is unrolled spatially on hardware, as presented in the most right column in Fig~\ref{fig:intro-table}. The tensors are compressed in $(A,B)$ order.

$\bm{U_{M}U_K,U_KU_{N}}$ \textbf{(GEMM, TPU-like)}.
Fig~\ref{fig:sub-accelerators}a shows an output stationary systolic array \cite{samajdar2020systematic}. Both input matrices are in dense (uncompressed) format and are fed into the Processing Element (PE) array in a store-and-forward manner. Each PE receives input data from the top and left side, computes a MAC operation, stores the partial sum locally, and forwards out data from the bottom and right side. Once all computations are finished, the completed partial sums are sent to the output accumulation buffer. With a flexible interconnect \cite{kwon2018maeri, extensor, geng2019awb}, it is possible to utilize M $\times$ N PEs in parallel.

$\bm{U_{M}C_K,U_KU_{N}}$ or $\bm{U_{M}U_K,U_{N}C_K}$ \textbf{(SpMM, EIE-like)}.
Based on the accelerator EIE \cite{eie}, Fig~\ref{fig:sub-accelerators}b shows a scenario in which Matrix \textbf{A} is in dense (uncompressed) format, and Matrix \textbf{B} is in $U_NC_K$ format. Each column of Matrix \textbf{B} ($row\_ids$ and $values$) is loaded into the buffer of a PE, hence it is possible to utilize a maximum of N PEs in parallel. Matrix A is streamed through a bus to all PEs. An index comparison module is used to find valid nonzero computations by using the bus position of Matrix \textbf{A} and $row\_ids$ of Matrix \textbf{B}. The valid computations are then scheduled onto the MAC Queue. Each PE has an output register to accumulate the partial sums. Note that the architecture also support a scenario in which matrix A is compressed in $U_MC_K$ and matrix B is uncompressed. 

$\bm{U_{M}C_K,U_{N}C_K}$ \textbf{(SpGEMM, ExTensor-like)}.
Based on the accelerator Extensor \cite{extensor} (inner-product), Fig~\ref{fig:sub-accelerators}c shows Matrix \textbf{A} in $U_MC_K$ format, and Matrix \textbf{B} in $U_NC_K$ format. Comparing each row of Matrix \textbf{A} with every column of Matrix \textbf{B}, Matrix \textbf{A}'s $col\_ids$ and \textbf{B}'s $row\_ids$ go into a hardware intersection unit that quickly finds matching indices. Using the indices, the corresponding $values$ are fetched from both matrices. The $values$ are then distributed through a Network-on-Chip (NoC) to the corresponding PEs for computation. When comparing each row of Matrix \textbf{A} with every column of Matrix \textbf{B}, the maximum PE utilization is N. When comparing each column of Matrix \textbf{B} with every row of Matrix \textbf{A}, the maximum PE utilization is M.

$\bm{U_KC_{M},U_KC_{N}}$ \textbf{(SpGEMM, OuterSPACE-like)}.
Based on the accelerators OuterSPACE \cite{outerspace} and SCNN \cite{parashar2017scnn} (outer-product), Fig~\ref{fig:sub-accelerators}d shows a scenario in which Matrix \textbf{A} is in $U_KC_M$ format, and Matrix \textbf{B} is in $U_KC_N$ format. Iterating over the K dimension \footnote{For hardware generation ease, we unrolled the K dimension spatially (shown in Fig~\ref{fig:taco-output-new}d), hence the maximum PE utilization is K.}, the accelerator fetches the indices ($row\_ids$, $col\_ids$) and nonzero data ($values$) for outer product computation. Each PE owns a partition of the output matrix accumulation, and a NoC sends the fetched input meta(data) to the corresponding PE. Necessary Matrix \textbf{B}'s (meta)data is first loaded into the PEs' B buffers, and then Matrix \textbf{A}'s (meta)data is streamed in. Each PE also has its own accumulation buffer that contains the partial sums of each output matrix position the PE owns. The indices ($row\_ids$, $col\_ids$) are used to read and write to the correct location.

$\bm{U_KC_{M},U_{N}C_K}$ \textbf{(SpGEMM, MatRaptor-like)}.
Based on the accelerator MatRaptor \cite{srivastava2020matraptor} (column-wise-product or Gustavson’s algorithm), Fig~\ref{fig:sub-accelerators}e shows a scenario in which Matrix \textbf{A} is in $U_KC_M$ format, and Matrix \textbf{B} is in $U_NC_K$ format. Iterating over Matrix \textbf{B}'s columns, a controller reads Matrix \textbf{A}'s columns that corresponds to Matrix \textbf{B}'s $row\_ids$. Each PE owns a column partition of the output matrix, hence the maximum PE utilization is N. Necessary Matrix \textbf{B}'s (meta)data is first loaded into the PEs' B buffers, and then Matrix \textbf{A}'s (meta)data is streamed in.  Matrix \textbf{A}'s $col$ is used to compare with Matrix \textbf{B}'s $row\_ids$ to schedule useful computations into the MAC Queue.
Matrix \textbf{A}'s $row\_ids$ are used to read and write to the correct location within the accumulation buffer. Note that Gamma \cite{zhang2021gamma} also uses Gustavson’s algorithm, although Gamma traverses row-wise through the matrices rather than column-wise.

\insertFigure{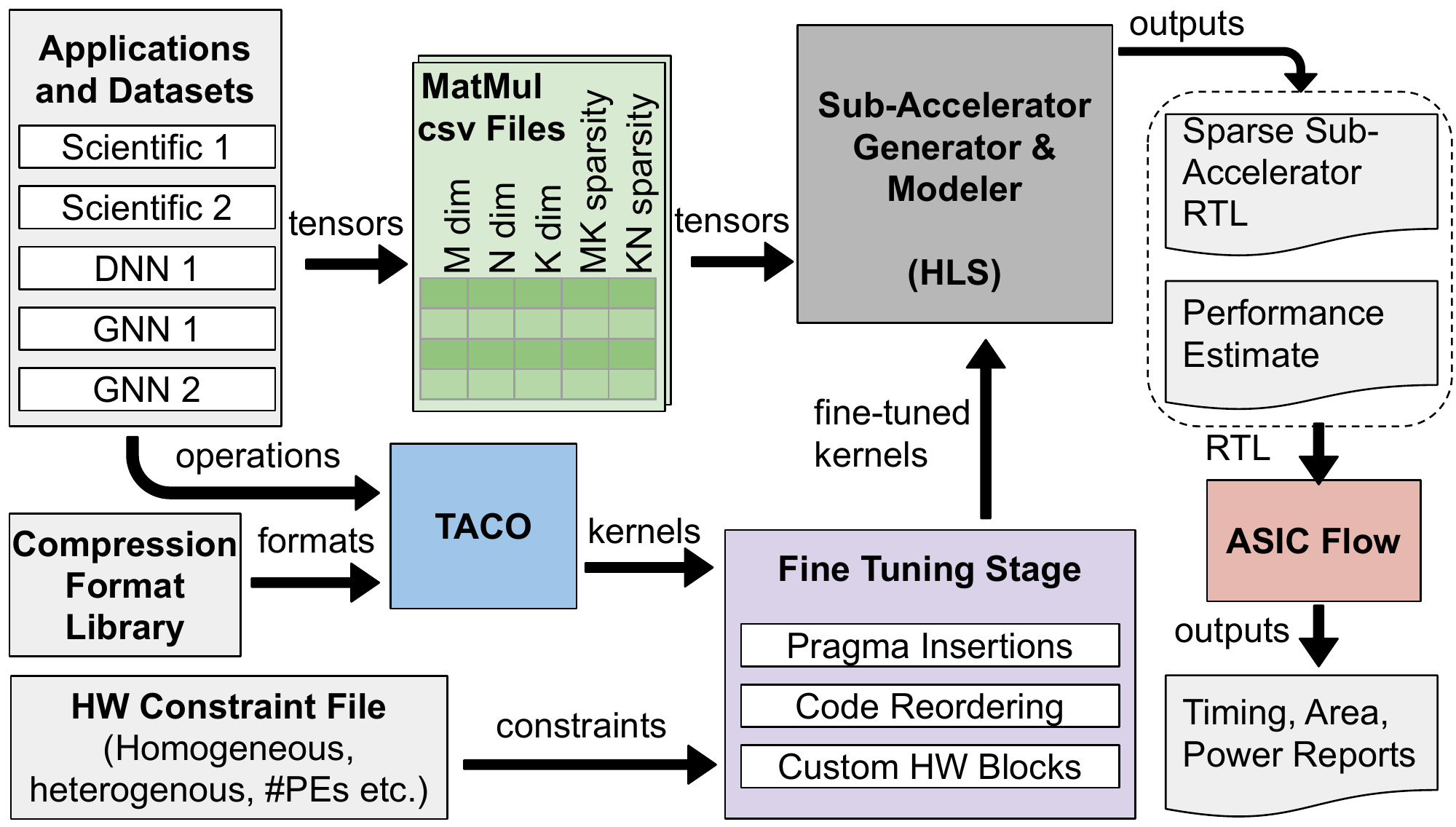}{\textit{Hard TACO} design flow overview for estimating sub-accelerator performance, power, and area cost.}

\subsection{Hard TACO Design Flow}
\label{sec:designflow}

\textit{Hard TACO} generates the sparse sub-accelerators in Section~\ref{sec:sparseclass} to get fast performance and hardware resource consumption for design space exploration. Fig~\ref{fig:design-overview-new} shows our design flow at a high level. First, matrix multiplication operations using different CCFs are sent as inputs into the TACO compiler. The TACO compiler generates C++ code to the `\textit{Fine Tuning Stage}' along with user defined hardware constraint parameters, which includes number of PEs, memory type, etc.

The `\textit{Fine Tuning Stage}' inserts pragmas and reorders the TACO output to meet the hardware constraints. Fig~\ref{fig:taco-output-new} shows where HLS UNROLL pragmas are inserted to unroll a loop spatially, which creates multiple hardware instances for parallel computations. The PIPELINE pragma attempts to achieve low initiation interval within a code block. BIND$\_$OP pragma hints operations to be synthesized using LUTs or DSPs. For this work, we use the BIND$\_$OP pragma to synthesize the accelerator PEs onto the DSPs. ALLOCATE pragma limits how many hardware instances can be generated of a particular compute function. ARRAY$\_$PARTITION pragma determines the local buffer bandwidth by modeling memory instances to behave like SRAMs or registers.

Next, the fine-tuned kernels go directly into the HLS tool. Many HLS tools translate code written in a high level language, such as C++, into RTL. For this work, we use Xilinx Vitis HLS, though we note that any other HLS tool could be used. The outputs include RTL, FPGA hardware consumption, and performance estimates. The RTL then goes through an ASIC flow to get more detailed timing, area, and power reports.

\textit{Hard TACO generates quick hardware cost estimates of various sparse sub-accelerators, which become building blocks for a larger heterogeneous accelerator design.}

\section{AESPA Architecture Template}
\label{sec:aespa}

This section proposes \aespa, a heterogeneous sparse accelerator. Fig~\ref{fig:aespa} shows the high level architecture template and specification for building \aespa. The die size is capped at 600 $mm^2$, which is approximately the same size as TPU v2 \cite{jouppi2020domain}. HBM memory size and bandwidth is set to 32 GB and 1 TB/second respectively. Global scratchpad size and bandwidth is set to 64 MB and 8.192 TB/second respectively. We utilize a highly flexible NoC, often used in previous works \cite{geng2019awb, kwon2018maeri, extensor, eyeriss2}, to send data from the global buffer to all PEs.

\subsection{Sub-Accelerator Cluster}

The conglomerate of sub-accelerator type, memory system, workload dimension, and sparsity ratio determines the operational intensity and performance of a given kernel. \aespa enables flexibility by having various sub-accelerator clusters (shown in Fig~\ref{fig:aespa}). The number of PEs in each sub-accelerator cluster is decided based on performance and energy metrics across a diverse workload set. It is a parameter for design space exploration in the evaluation section (Section~\ref{sec:eval}). Using the accelerator template, we create many possible configurations of \aespa{s}.

The number of PEs also depend on the available area for compute allocated on the die. After accounting for the memory and peripheral logic area, only 202.96 $mm^2$ of space is left for compute. Depending the type and quantity of sub-accelerators allocated, the peak TFLOPS/second ranges from 9.98 to 34.56.

\subsection{Flexible Global Buffer}
\aespa consists of a double buffered and flexible Global scratchpad which can support flexible-sized partitions for different matrices and sub-matrices (Section~\ref{sec:scheduling-single}). The Global scratchpad is distributed such that each sub-accelerator is backed by one slice of the scratchpad.
The Global scratchpad can store matrix and sub-matrix data and meta-data in multiple layouts and multiple compression formats as required by the sub-accelerators. For example, Outer-space like sub-accelerator requires both matrices in a K-major layout with M and N ranks compressed while Extensor-like sub-accelerator requires matrix A in the M-major layout and matrix B in the N-major layout, both matrices having K rank compressed.


\insertFigure{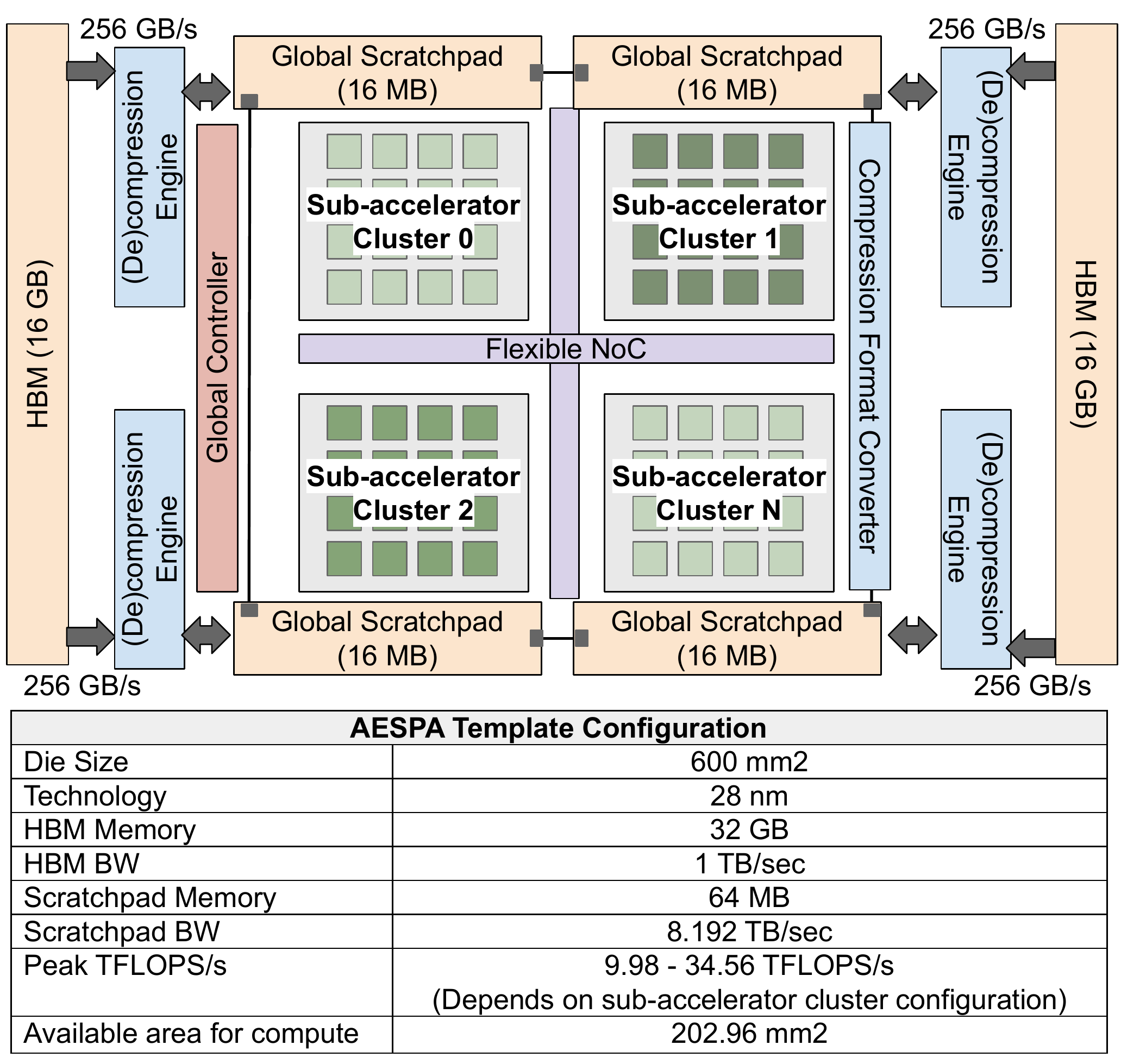}{\aespa architecture template diagram and specifications.}

\insertWideFigure{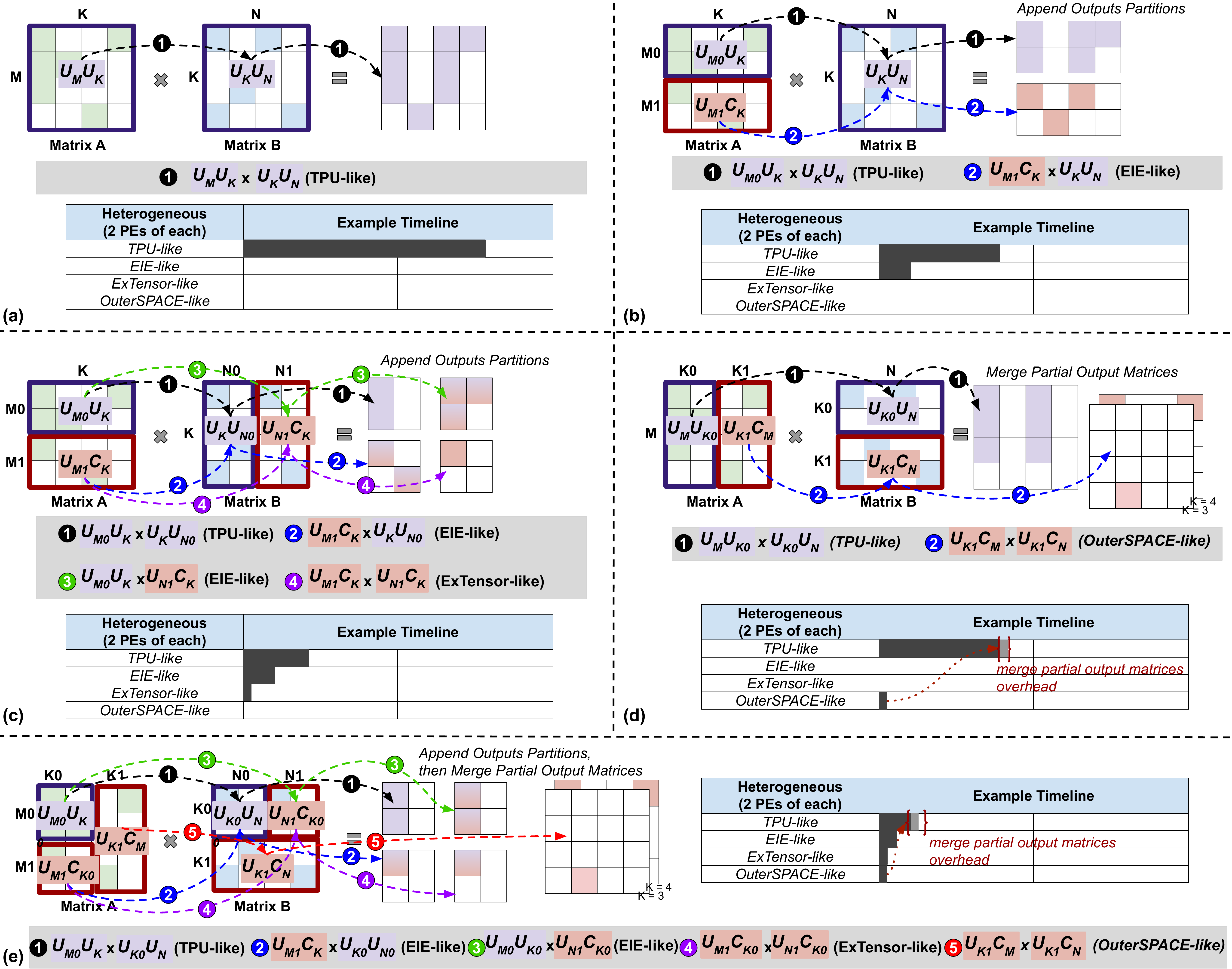}{Examples showing how a single matrix multiplication kernel can be scheduled onto heterogeneous sparse accelerator to achieve peak performance on a single kernel. (a) Scenario when only TPU-like sub-accelerator is active. (b) Scenario when TPU and EIE-like sub-accelerators are active. (c) Scenario when TPU, EIE, and ExTensor-like sub-accelerators are active. (d) Scenario when TPU and OuterSPACE-like sub-accelerators are active. (e) Scenario when TPU, EIE, ExTensor, and OuterSPACE-like sub-accelerators are active. Final runtime is the maximum runtime across sub-accelerators. \vspace{3mm}}

\subsection{(De)compressor and Format Conversions}
\label{sec:converter}
Decompressors are used so that the data in HBM can be stored in a format with low memory footprint \cite{rhu2018compressing}. This feature allows faster transfer time and better energy efficiency, as a word of memory transfer from main memory is $\sim$6400$\times$ greater in energy consumption than a single integer add operation \cite{eie}. It is possible to bypass decompression if the sparse sub-accelerator needs to compute on the compressed format directly. Additionally, format converters are used to enable sub-accelerators under a scenario in which the data transferred from host is compressed in a format that does not match the accelerator's CCF. For example, a $U_KC_M,U_KC_N$ based accelerator is instantiated, but the input tensor is transferred from host to accelerator as U$_M$C$_K$; therefore, a U$_M$C$_K$ $\longrightarrow$ U$_K$C$_M$ hardware block next to the accelerator is required. A typical workaround is to convert the tensors to the right format in the host before transferring, but at the cost of potentially increasing the transfer size. Recent works propose hardware units that do format conversion on the fly \cite{asgari2020copernicus}. The \aespa template determines the type and quantity of sub-accelerators used to create a large heterogeneous sparse accelerator. Different \aespa configurations are explored in the evaluation section.

\section{Scheduling for AESPA}
\label{sec:scheduling}

Scheduling matrix multiplication kernels onto sparse accelerators has its own set of problems. Input tensors are expected to be compressed in a specific format for computation. For example, a $U_MU_K,U_NC_K$ sub-accelerator requires Matrix \textbf{A} to be in $U_MU_K$ and Matrix \textbf{B} to be in $U_NC_K$. Homogeneous accelerators contain PEs that can utilize a specific compute format combination. Heterogeneous accelerators, on the hand, contains different PEs with unique sparse controller and microarchitectures. If the entire tensor matrix is compressed in one format, a portion of the PEs will remain underutilized. 

The first scheduling technique (discussed in Section~\ref{sec:scheduling-single}) partitions the input tensors into different compression formats to maximize compute utilization of all sub-accelerators within a heterogeneous accelerator. This maximizes peak performance of a single kernel for latency critical tasks. 

The second scheduling technique (discussed in Section~\ref{sec:scheduling-partition}) allocates many independent kernels onto to the heterogeneous accelerator in parallel. This is similar in concept to multi-tenancy proposed in recent works \cite{prema, planaria}.



\subsection{Single Kernel Scheduling (Max TFLOPs) Example}
\label{sec:scheduling-single}
Fig~\ref{fig:split-format-new-extended} shows the benefit of \textit{single kernel scheduling} on a single matrix multiplication operation. The examples model a heterogeneous accelerator with four sub-accelerators: $U_MU_K,U_KU_N$ (TPU-like), $U_MU_K,U_NC_K$ \& $U_MC_K,U_KU_N$ (EIE-like), $U_MC_K,U_NC_K$ (ExTensor-like), and $U_KC_M,U_KC_N$ (OuterSPACE-like). Each sub-accelerator has two PEs each for a total of 8 PEs. This example assumes that the system is compute bounded, and that there is sufficient memory bandwidth.

Fig~\ref{fig:split-format-new-extended}a shows a baseline case of $U_MU_K,U_KU_N$. Only the TPU-like sub-accelerator can compute this CCF combination. As a result, 6 of the total 8 PEs remain idle. The runtime is approximated by the number of iterations from the nested loop shown in Fig~\ref{fig:taco-output-new}a. In the case of Fig~\ref{fig:split-format-new-extended}a, given that the M, N, K dimensions are all four, the total number of iterations is 64. With two parallel PEs for this TPU-like sub-accelerator, the total execution is 32 cycles.

Fig~\ref{fig:split-format-new-extended}b shows an example with Matrix \textbf{A} split in half across the M dimension. The top half is uncompressed ($U_{M0}U_K$) and the bottom half is compressed in $U_{M1}C_K$. The hybrid format can be preprocessed beforehand, or converted with the hardware format converters found in \aespa. Both TPU and EIE like sub-accelerators are active. Note that the number of iterations for sparse sub-accelerators depends on the sparsity of the input tensor. In Fig~\ref{fig:split-format-new-extended}b, there is approximately one nonzero element per row of $U_{M1}C_K$, which determines the loop count of line 22 of Fig~\ref{fig:taco-output-new}. The TPU-like runtime is half of the previous example because the M dimension is split in half. The EIE-like runtime is 4 cycles, calculated by M1 $\times$ K $\times$ N $\times$ MK density divided by number of PEs.



Fig~\ref{fig:split-format-new-extended}c has Matrix \textbf{A} split across the M dimension and Matrix \textbf{B} split across the N dimension. This scenario activates three out of the four sub-accelerators (TPU-like, EIE-like and ExTensor-like). The TPU-like sub-accelerator cluster has a runtime of 8 cycles, from M0 $\times$ K $\times$ N0 / number of PEs. The EIE-like accelerator is utilized for two computation parts, highlighted in Fig~\ref{fig:split-format-new-extended}c computation \circled{2} and \circled{3}. Part \circled{2} runtime is half of Fig~\ref{fig:split-format-new-extended}b's, as the N dimension is split by half. Part \circled{3} runtime is similar to \circled{2}. The total computation cycle for EIE-like is 4 cycles. ExTensor-like runtime is a cycle, approximated by M1 $\times$ K $\times$ N1 $\times$ MK density  $\times$ KN density divided by number of PEs.

Fig~\ref{fig:split-format-new-extended}d is split across the K dimension for both input matrices. The TPU-like runtime is 16 cycles, from M $\times$  K0 $\times$ N / number of PEs, while the OuterSPACE-like runtime is a cycle, approximated by M $\times$ K1 $\times$ N $\times$ MK density  $\times$ KN density divided by number of PEs. Since the reduction dimension (K) is split, remaining partial output matrices are merged at the end.


Fig~\ref{fig:split-format-new-extended}e is split across the M, N, and K dimension. This allows all of the sub-accelerator types to compute a single kernel together. The TPU-like sub-accelerator cluster has a runtime of 4 cycles, from M0 $\times$ K0 $\times$ N0 / number of PEs. The EIE-like accelerator is utilized for two computation parts, highlighted in Fig~\ref{fig:split-format-new-extended}e computation \circled{2} and \circled{3}. The runtime for EIE-like is 2 cycles (1 cycle for \circled{2}, and 1 cycle for \circled{3}). ExTensor-like runtime is a cycle, approximated by M1 $\times$ K0 $\times$ N1 $\times$ MK density  $\times$ KN density divided by number of PEs. OuterSPACE-like runtime is a cycle, similar to that of Fig~\ref{fig:split-format-new-extended}d. Since the reduction dimension (K) is split, remaining partial output matrices are merged at the end.

\textit{Our single kernel scheduling strategy for heterogeneous sparse accelerators enables high utilization for all available PEs of various sub-accelerator types. As shown in Fig~\ref{fig:split-format-new-extended}e, the strategy improves performance, which is important for latency critical kernels. The performance benefit depends on the \aespa configuration, memory bandwidth, and workload. (Discussed in more detail in (Section~\ref{sec:eval}).}

\insertFigure{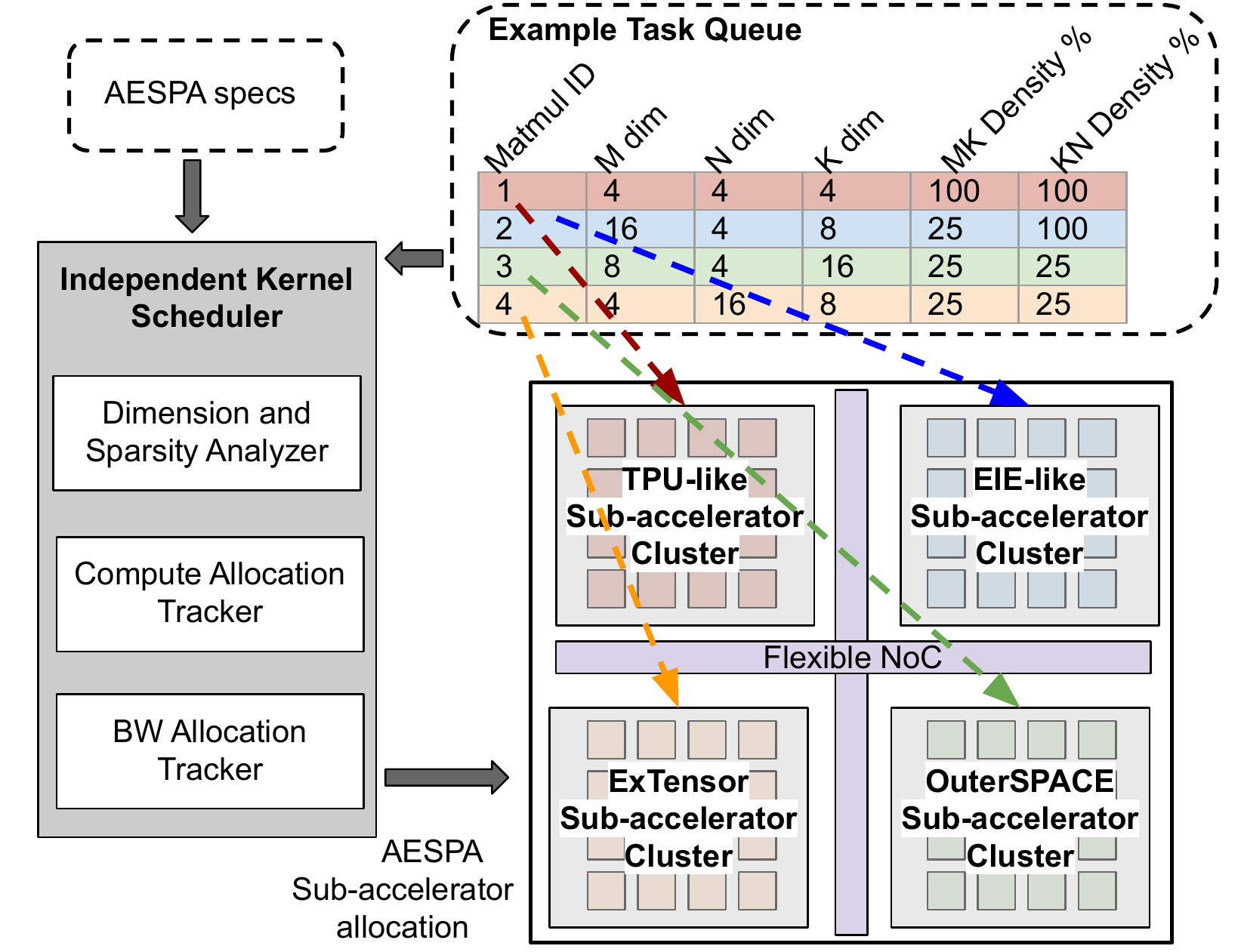}{High level example showing how different workloads are allocated to different sub-accelerators in a heterogeneous sparse accelerator.}

\subsection{Many Kernels Scheduling Example}
\label{sec:scheduling-partition}
For large datacenters, it is common to have many kernels in a queue waiting to be executed. Rather than partitioning the tensors into hierarchical formats to achieve high TFLOPS/second per kernel as observed in the \textit{`single kernel scheduling'}, \textit{`many kernels scheduling'} optimizes for multiple kernels, each with tensors compressed in one format. Fig~\ref{fig:partitioning-aespa} shows a high level figure with four matrix multiplication kernels in the task queue. The figure also shows an \aespa configuration of four sub-accelerator types (TPU, EIE, ExTensor, and OuterSPACE-like) with 16 PEs each. 

The first task (red) is ideal for the TPU-like sub-accelerator. This is because the tensors are completely dense, and all 16 PEs can be utilized from the M $\times$ N bound (refer to the parallelism dimension bound table in Fig~\ref{fig:intro-table}). The second task (blue) is ideal for the EIE-like sub-accelerator. This is because the M dimension can be fully unrolled onto the PEs, and there is one input matrix that is relatively sparse. The third (green) and fourth (orange) tasks have two matrices that are relatively sparse, so they run more efficiently on sub-accelerators that support SpGEMM. The third (green) has a K dimension that can fully unrolled onto an OuterSPACE-like sub-accelerator, and the fourth (orange) has a N dimension that can be fully unrolled onto an ExTensor-like sub-accelerator.

\textit{Our many kernel scheduling strategy enables good utilization for multiple kernels running in parallel. However, this again depends heavily on the memory bandwidth, workload, and \aespa configuration. (Discussed in more detail in (Section~\ref{sec:eval}).}

\begin{table}[t]\footnotesize\centering
\setlength\tabcolsep{4 pt}
\caption{\small Tensor characteristics found in various applications.}
\begin{tabular}{lllll}
\toprule
\textbf{Name} & \textbf{Application} & \textbf{Dimension} & \textbf{Density  \% }   \\
\textbf & \textbf & \textbf{(M,K,N)} & \textbf{(MK,KN)}   \\
\toprule
    chem97ZtZ & Stat Problem & 2.5k$\times$2.5k$\times$1.2k & 0.11,100   \\
    journals  & Weighted Graph & 124$\times$124$\times$62 & 78.5,100   \\ 
    m3plates  & Acoustics & 11k$\times$11k$\times$5.5k & 0.0054,100  \\ 
    synthetic$\_$dense  & Varies  & 5k$\times$5k$\times$2.5k & 100,100  \\ 
    bibd$\_$81$\_$3  & Combinatorial  & 3.2k$\times$85k$\times$43k & 0.093,100  \\ 
    speech  & Deep Learning  & 7.7k$\times$2.6k$\times$1.3k & 5,100  \\ 
    gnmt  & Deep Learning & 1.6k$\times$1k$\times$36k & 50,30 \\
    transformer & Deep Learning &  32k$\times$84$\times$1k & 50,30  \\
    citeseer  & GNN & 3.3k$\times$3.3k$\times$3.7k & 0.11,0.85 \\
    
\bottomrule
\end{tabular}
\label{table:workload}
\end{table}

\section{Methodology}

We first model each basic sub-accelerator class (refer to Sec~\ref{sec:sparseclass}) using the \textit{Hard TACO} design flow. For HLS tool, we used Xilinx Vitis and ran hardware emulation on Alveo U50. Functional correctness was verified for all sub-accelerators. We get the RTL, estimated FPGA hardware resource consumption and performance from Vitis. The generated RTL then goes through an ASIC flow using 28nm technology for a more detailed area, power, and timing report. Synopsys DC compiler was used for synthesis and Cadence Innovus was used for place and route. 

The hardware generation time on Vitis was significant when modeling large designs (128 PEs +), so we designed small sub-accelerator cores, each utilizing 16 int32 PEs. Each core was also implemented to have a local buffer, which stores three 16 $\times$ 16 tiles (one for each tensor operand). The local buffer size is different for sparse sub-accelerators as they require more metadata storage. Floating point units were not enabled when generating the sub-accelerators from Vitis. To approximate the cost of enabling floating point operations, we add the ASIC area and power costs of our internal FP units to the post-ASIC flow sub-accelerator reports. We scale the area and power of the generated small sub-accelerator units to larger designs (4196 PEs +) linearly. Though not realistic, we believe that it is still a fair estimation, given each generated sub-accelerator is self sufficient on its own (contains its own local buffers and controllers). With the generated sub-accelerators, we are able to approximate the amount of TFLOPS/s achievable given an area constraint.

To generalize how an architecture will behave for realistic workloads (HPC \cite{suitesparse}, DNN \cite{baidudeepbench} , and GNN \cite{kersting2016benchmark, yanardag2015deep}) shown in Table~\ref{table:workload}, we developed an analytical model that approximates the performance by first estimating the tripcount of the compute loop of each matrix multiplication kernel shown in Fig~\ref{fig:taco-output-new}. The number of iterations depend on the tensor dimension and sparsity. We assume uniform random sparsity. The memory bandwidth of the \aespa system is also integrated into the model, as sparse computations are often memory bounded \cite{srivastava2020matraptor,extensor,zhang2020sparch}. For energy cost, we consider the utilization of the accelerator and the on-chip data movement. 


To schedule operations onto heterogeneous accelerators, we conduct a search to find the best way to partition/configure a certain operation derived from our proposed strategies. For simplicity, we assume that the data compressed in the desired format(s) is sent directly from the host. Different configurations of \aespa are evaluated against state-of-the art accelerators presented in Fig~\ref{fig:intro-table}.

\section{Evaluation}
\label{sec:eval}

\insertFigure{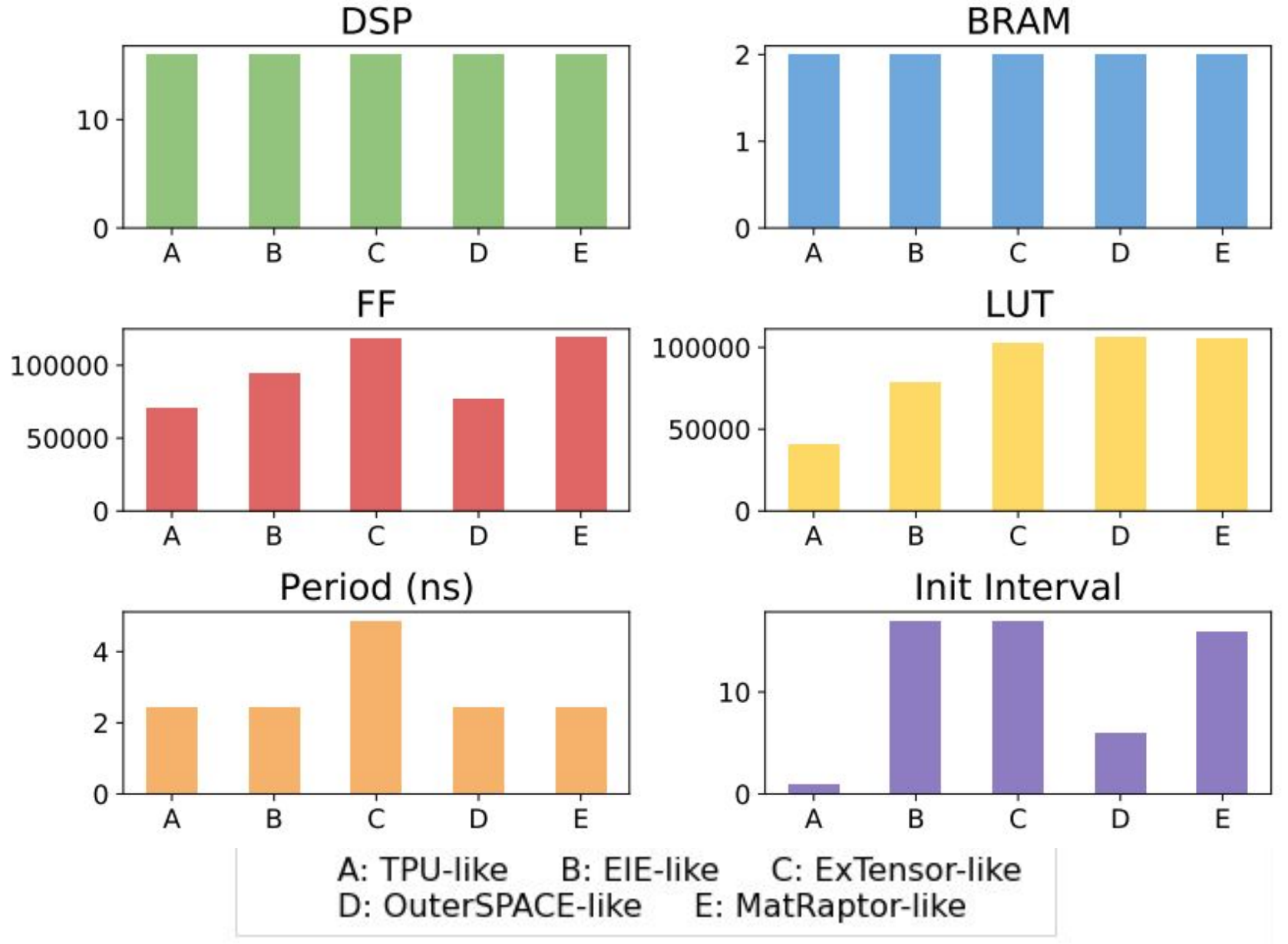}{Xilinx U50 FPGA hardware resource and performance report on different sub-accelerator types with 16 PEs each.}

\insertFigure{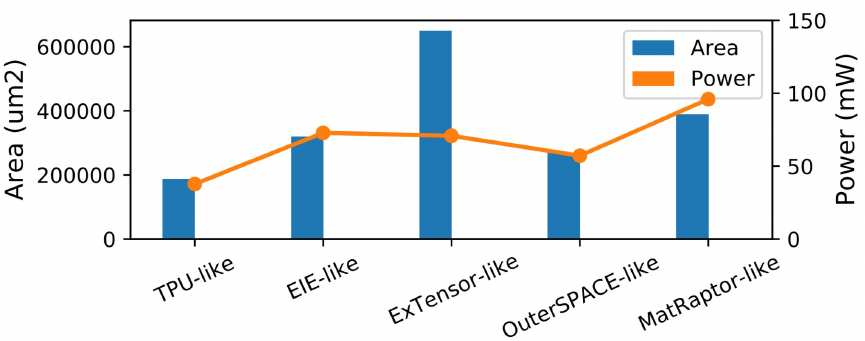}{28nm ASIC area and power report on different sub-accelerator types with 16 PEs each.}

This section first analyzes the sub-accelerator costs using \hardtaco. Then, we evaluate the performance of \aespa with our scheduling techniques against state of the art accelerator designs.

\subsection{Hard TACO Reports}
Fig~\ref{fig:u50} shows the Xilinx U50 FPGA results on different sub-accelerators. There are 16 PEs in each HLS design, hence why there are 16 DSPs. The number of FFs is high for ExTensor and MatRaptor-like sub-accelerators, while the number of LUTs is high for ExTensor, OuterSPACE, and MatRaptor-like sub-accelerators. TPU-like design has the least number of FFs and LUTs. This is intuitive because it does not need any extra metadata controller or index indirections that are required in sparse sub-accelerators. This TPU-like property is confirmed by the initiation interval of 1. An initiation interval of 1 means that it is completely pipelined, while a high initiation interval will induce a lot of stalls. EIE, Extensor, and MatRaptor-like all have initiation intervals of 17, 17, and 16 respectively. OuterSPACE-like design has a relatively lower initiation interval of 6. The period of ExTensor-like on the FPGA is almost 2$\times$ greater than the others. The inner product dataflow that ExTensor implements contain lots of metadata intersections, which may not be efficient on the FPGA fabric.

To gain better understanding on the sub-accelerator designs, we synthesize the generated RTL on a 28nm process. Fig~\ref{fig:asic} shows the area and power report. All sub-accelerators met timing at 1 GHz. Floating point units overhead are added on top. Additionally, 32 32-bit wide FIFOs of depth 10 are added per initiation interval. This is a rough estimate to predict what the sub-accelerator overhead would be if it is fully pipelined with custom ASIC implementations. ExTensor-like's area is the largest, with almost 3$\times$ greater than a TPU-like design. MatRaptor-like design is the most power-hungry among all, while OuterSPACE-like design is relatively low in area and power. We emphasize that custom hand-designed ASIC implementations may differ from the results gathered from the HLS tool, though we believe they are fair estimates of the building blocks for design space exploration. We utilize the area overhead of each sub-accelerator to find the peak TFLOPS/second of various accelerator types shown in Fig~\ref{fig:intro-table}.




\insertWideFigure{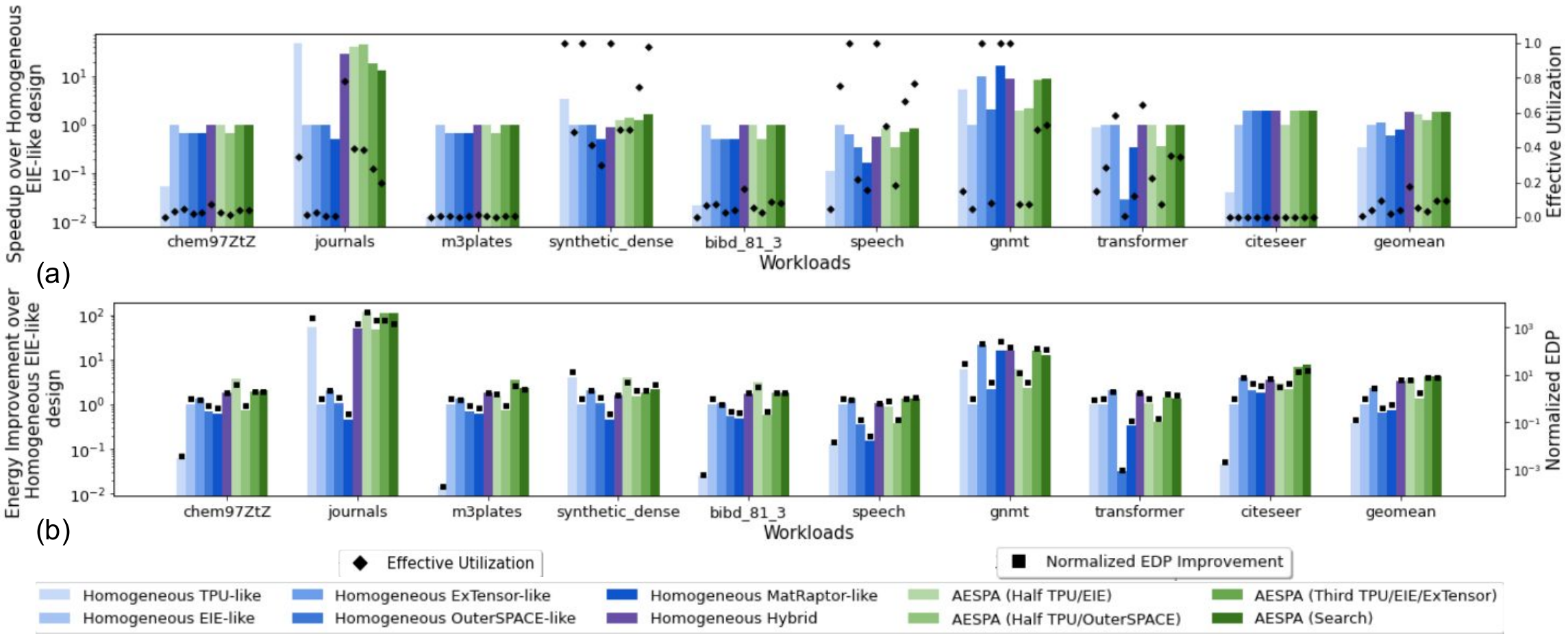}{Evaluation against Homogeneous EIE-like with limited memory bandwidth. (a) shows speedup and effective utilization. (b) shows normalized energy and EDP improvement. \vspace{5mm}}

\insertWideFigure{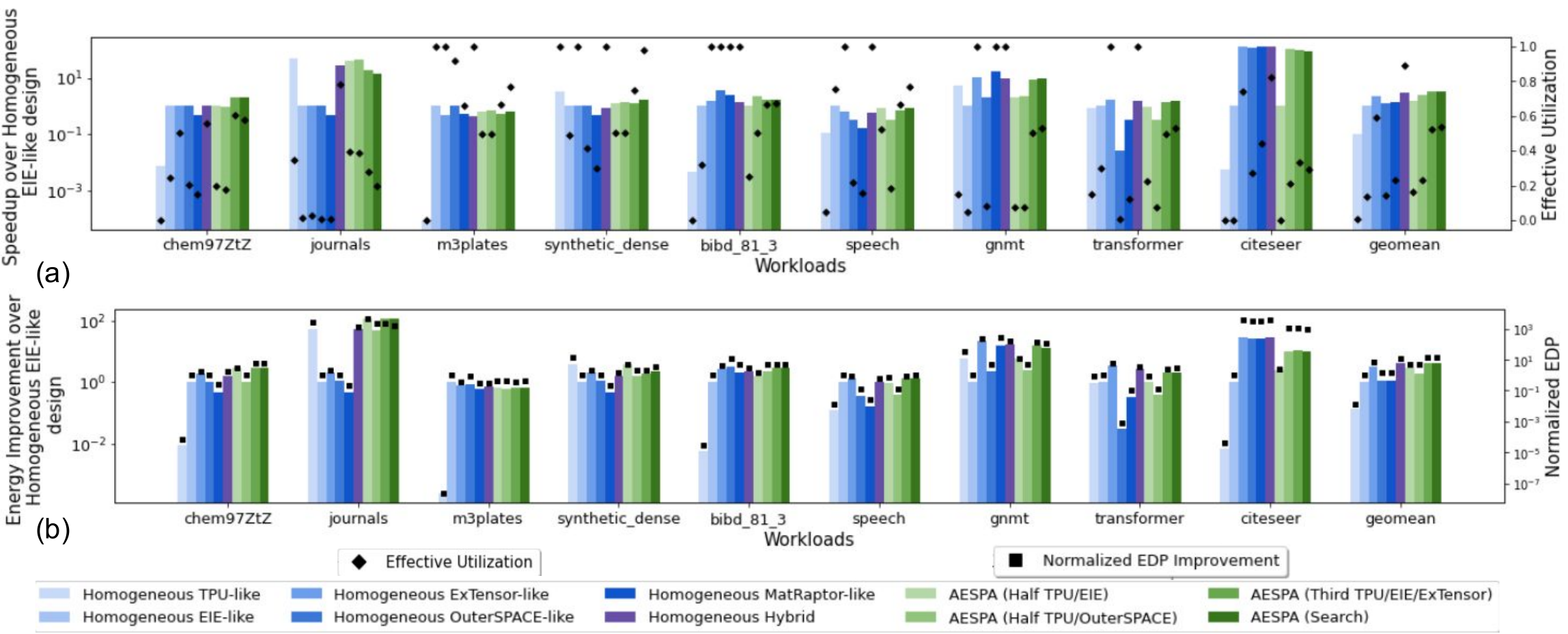}{Evaluation against Homogeneous EIE-like with unlimited memory bandwidth. (a) shows speedup and effective utilization. (b) shows normalized energy and EDP improvement. \vspace{5mm}}

\subsection{\aespa Single Kernel Scheduling Evaluations}
Fig~\ref{fig:limited-bw} and Fig~\ref{fig:unlimited-bw} shows how \aespa compares to other baselines when computing individual matrix multiplications. The first six baselines are homogeneous accelerators. A homogeneous accelerator only contain one type of sub-accelerator. We also compare against a homogeneous hybrid design, in which the sub-accelerator can support the dataflows of all TPU-like, EIE-like, and ExTensor-like. The cost of this flexibility is extra area and power, which then reduce the peak TFLOPS/s achievable under an area constraint. Fig~\ref{fig:intro-table} shows that the TFLOPS/s of a homogeneous hybrid is just 8.96. For heterogeneous designs (\aespa), it is possible to achieve flexibility while still using basic sub-accelerators that can only support one type of sparse dataflow.

Fig~\ref{fig:limited-bw} and Fig~\ref{fig:unlimited-bw} show performance graphs with 1 TB/s and unlimited memory bandwidth respectively. Both figures Part A shows the speedup over Homogeneous EIE-like design on the left Y axis and effective utilization on the right Y axis. Effective utilization is the percentage of effectual work done by the accelerator. It depends on (1) the parallelism dimension bound and the sparsity support the sub-accelerator enables (shown in Fig~\ref{fig:intro-table}). Part B left Y axis shows the energy efficiency improvements over a Homogeneous EIE-like design, and the right Y axis shows the normalized EDP improvement.

M3plates (refer to Table~\ref{table:workload}) has a very sparse Matrix \textbf{A}. The operational intensity is low and gets limited by the memory bandwidth, as shown by the low utilization points across all designs in Fig~\ref{fig:limited-bw}a. Fig~\ref{fig:unlimited-bw}a shows that with unlimited bandwidth, the performance utilization points increase drastically. Homogeneous TPU-like's utilization in this case is still low because it does not have any sparsity support. Citeseer also follows a similar pattern. Fig~\ref{fig:limited-bw}a shows significant performance degradation for Homogeneous OuterSPACE-like on Transformer. This is because the K dimension for the workload is small (value of 84), and OuterSPACE-like is generated with the K dimensions unrolled spatially; therefore, the utilization is bounded by the dimension of the workload. \aespa is able to have variable parallelism bounds, hence why \aespa (Half TPU/OuterSPACE) is able to achieve higher utilization. We compared against four configurations of \aespa, the first three have fixed ratios of sub-accelerators, while the last is high performance configuration searched by our model. Homogeneous hybrid has the highest effective utilization, for both limited and unlimited bandwidth. However, the smaller peak TFLOPS/s prevent it from obtaining larger speedup. 

\aespa with our single kernel scheduling strategy is able to achieve 1.96$\times$ speedup and 7.9$\times$ EDP geomean improvement over Homogeneous EIE-like at a 1 TB/s memory bandwidth. With unlimited bandwidth, \aespa is able to achieve 3.3$\times$ speedup and 14.1$\times$ EDP geomean improvement. Against a Homogeneous Hybrid design, \aespa achieves 1.03$\times$ speedup and 1.28$\times$ EDP geomean improvement. With unlimited bandwidth, it becomes 1.13$\times$ speedup and 1.20$\times$ EDP geomean improvement.

\subsection{\aespa Many Kernel Scheduling Insights}

Rather than achieving peak performance for a single kernel, as shown in the previous section, it is also possible to run multiple kernels in parallel on \aespa. The workload characteristics and resource availability determine which kernel gets mapped to which sub-accelerator cluster. In Fig~\ref{fig:many-kernel}, we evaluated all designs with unlimited bandwidth. \aespa is able to stay within 6\% of the best baseline configuration for total runtime. With limited bandwidth, we observe more memory contention across parallel kernels, significantly impacting performance. We note that this analysis is very dependent on the workloads given, but it shows that there are many valid ways to run applications on a heterogeneous sparse accelerator.

\insertFigure{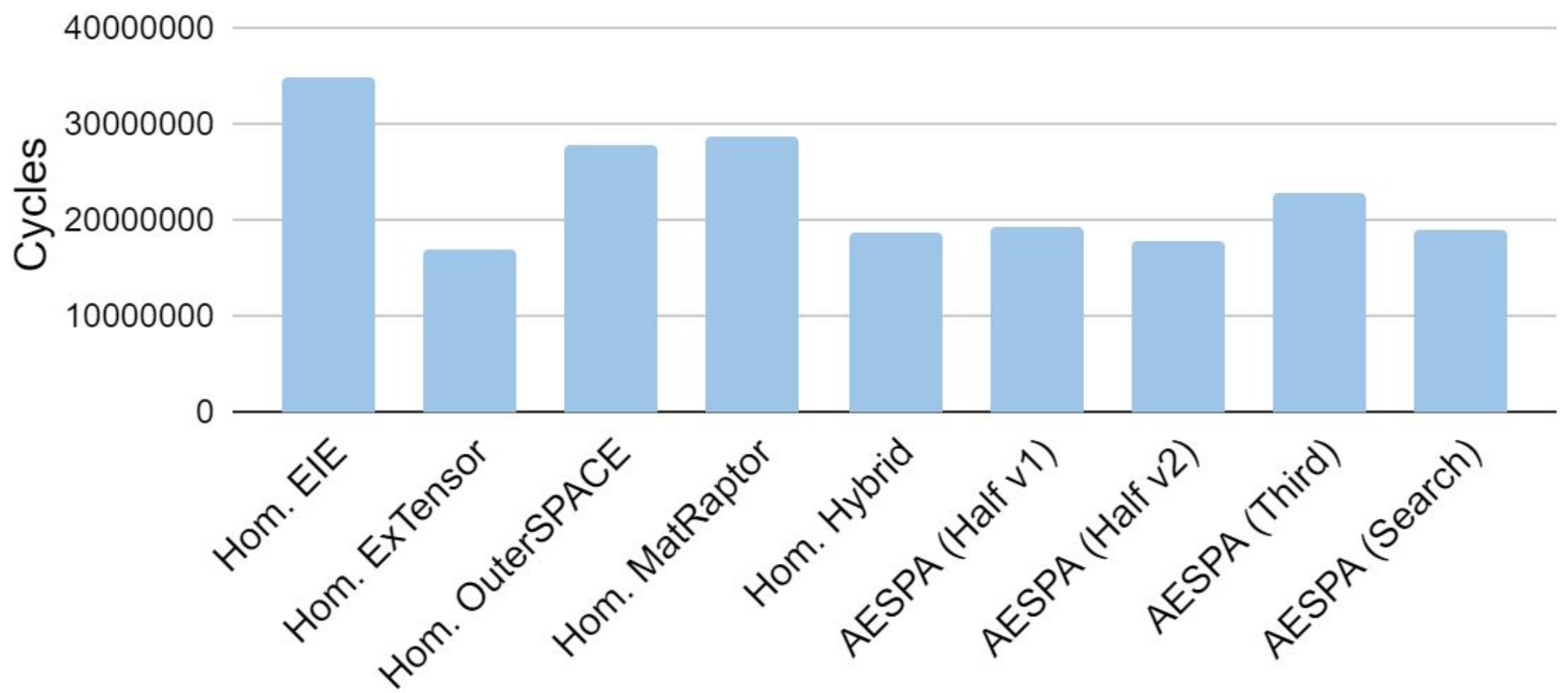}{Total cycles for completing all workloads using many kernel scheduling.}
\section{Related Work}

\textbf{Pre-RTL accelerator DSE}: Aladdin is a pre-RTL, power-performance accelerator simulator \cite{shao2014aladdin}. Aladdin is an alternate methodology to \hardtaco to quickly get hardware and performance estimates.
Interstellar~\cite{yang2020interstellar}, MAESTRO~\cite{kwon2019understanding} and Timeloop~\cite{timeloop} are analytical performance models for DNN accelerators.  Herald~\cite{herald} is an optimization framework for heterogeneous accelerator substrates for dense DNN workloads. It finds the optimal resource allocation for multiple-sub accelerators and determines the optimal schedule of multi-DNN workloads on the dense heterogeneous substrate. However, Herald only looked into dense accelerators, while we expand it to sparse accelerators. Numerous other works look into exploring parallelism for popular workloads \cite{song2019hypar, yan2020hygcn, hwang2020centaur, chen2021rubik, du2015shidiannao}.

\textbf{Accelerator RTL Generators.}
Spiral is a HLS wrapper for FFTs \cite{franchetti2018spiral} and MAGNet is a modular accelerator generator for neural networks \cite{venkatesan2019magnet}. DeepBurning-GL is a framework that generates different accelerators for GNNs. \cite{liang2020deepburning}. \textit{Hard TACO} takes a unique approach by utilizing the output of an established sparse tensor compiler \cite{tacoonline} to generate RTL. This allows people with compiler/algorithms background to intuitively understand what each sparse sub-accelerator is doing and their differences. 

\textbf{Sparse accelerators}: Sparsity has been a key optimization target for HPC and AI workloads. Numerous accelerators have been proposed for SpMM and SpGEMM acceleration~\cite{eie,outerspace,extensor,srivastava2020matraptor,eyeriss2, lee2018stitch, gondimalla2019sparten, sigma, kanellopoulos2019smash, yan2020hygcn,geng2019awb,liang2020engn,liang2020deepburning,Auten2020,gcnax,shi2021versagnn, srivastava2020tensaurus, sparseadapt, gospa, zhang2016cambricon}. Though there are a lot of unique sparse accelerators, we believe that they can all be grouped into classes based on their computation dataflow.

Recent works look into co-designing for sparse architectures. Sanger prunes the attention matrix for its reconfigurable architecture to exploit \cite{sanger}. ESCALATE utilized kernel decomposition to accelerate CNN models \cite{escalate}. G-Cos does a co-search on both the GNN and the accelerator to maximize accuracy and hardware efficiency. The authors also looked into sub-accelerator designs with various sparse dataflows \cite{zhang2021gcos}. SpAtten does pruning and quantization to achieve high energy efficiency \cite{wang2021spatten}.
%
%



\section{Future Works and Discussion}

There is a wide range of sparse accelerators presented in industry and academia, each with some unique optimizations that cannot be modeled efficiently with HLS. These optimizations include load balancing mechanisms, structured sparsity support, etc. Although HLS is perfect for fast architecture exploration, it is more realistic to build custom IPs for deployment. Additionally, the field of scheduling on heterogeneous sparse accelerators is relatively unexplored compared to its dense counterpart. There are numerous opportunities for new hierarchical compression formats to utilize different sparse sub-accelerators.

\bibliographystyle{IEEEtranS}
\bibliography{references}

\end{document}